\documentclass{llncs}

\usepackage{graphicx}
\usepackage{dsfont}        
\usepackage{exscale}       
\usepackage{amsmath}       
\usepackage{amssymb}
\usepackage{color} 
\usepackage{hyperref} 
\usepackage[all]{hypcap} 
\definecolor{darkred}{rgb}{0.5,0,0}
\definecolor{darkgreen}{rgb}{0,0.5,0}
\definecolor{darkblue}{rgb}{0,0,0.5}

\hypersetup{
linktocpage,
bookmarksnumbered=true,
colorlinks,
linkcolor=darkblue,
filecolor=darkgreen,
urlcolor=darkred,
citecolor=darkblue,
breaklinks=true,
pdfauthor={Chuanlong Li (chuanlongli@nuist.edu.cn)},
pdftitle={Stroke Lesion Segmentation with Visual Cortex Anatomy Alike Neural Nets},
pdfsubject={},
raiselinks,
}

\usepackage{booktabs}  

\usepackage{makeidx}  
\begin{document}
\frontmatter          
\pagestyle{headings}  

\mainmatter              
\title{Stroke Lesion Segmentation with Visual Cortex Anatomy Alike Neural Nets}
\titlerunning{Stroke Lesion Segmentation with Visual Cortex Anatomy Alike Neural Nets}  
%
\author{Chuanlong Li}

%
\authorrunning{Chuanlong Li} 
%
%
\institute{School of Electronic and Information Engineering, Nanjing University of Information Science \& Technology, Nanjing, China\\
\email{chuanlongli@nuist.edu.cn}
}

\maketitle              

\begin{abstract}
Cerebrovascular accident, or commonly known as stroke, is an acute disease with extreme impact on patients and healthcare systems and is the second largest cause of death worldwide. Fast and precise stroke lesion detection and location is an extreme important process with regards to stroke diagnosis, treatment, and prognosis. Except from the manual segmentation approach, machine learning based segmentation methods are the most promising ones when considering efficiency and accuracy, and convolutional neural network based models are the first of its kind. However, most of these neural network models do not really align with the brain anatomical structures. Intuitively, this work presents a more brain alike model which mimics the anatomical structure of the human visual cortex. Through the preliminary experiments on the stroke lesion segmentation task, the proposed model is found to be able to perform equally well or better to the de-facto standard U-Net. Part of the implementation will be made available at \href{https://github.com/DarkoBomer/VCA-Net}{https://github.com/DarkoBomer/VCA-Net}.
\end{abstract}

\section{Introduction}

According to the World Health Organization (WHO), cerebrovascular accident is a world disease with extreme impact on patients and healthcare systems and is the second largest cause of death worldwide \cite{leslie2020towards}\cite{lee2017deep}. Fast and precise stroke lesion detection and segmentation is an extreme important process with regards to stroke diagnosis, treatment, and prognosis. Except from the manual segmentation approach, machine learning based segmentation methods are the most promising ones when considering efficiency and accuracy, and convolutional neural network based models are the first of its kind \cite{scalzo2015data}\cite{vincent2015detection}\cite{stier2015deep}\cite{tan2020deepbrainseg}\cite{yu2019lstm}\cite{long2015fully}. However, these neural network models do not really align with the brain anatomical structures thus lack of explanatory characteristics of the model outcomes. This work draws inspiration from neuroscience world and presents a brain alike model which mimics the anatomical structure of the human visual cortex, aiming to facilitate the stroke diagnosis and improve the healthcare systems by giving a more intuitive and explainable computational model.

\section{Methodology}

U-Net has been the mostly adopted base network for most of the deep learning methods. The U-shape with skip connections is able to capture both low and high level features and has become the de-facto standard for most deep learning based segmentation networks \cite{ronneberger2015u}. Many stroke lesion segmentation networks has been proposed since then \cite{seo2020machine}\cite{mondal2018few}\cite{sharique2019parallel}\cite{stier2015deep}\cite{zhao2019automatic}\cite{liu2019deep}\cite{ho2019machine}\cite{dolz2018dense}\cite{valverde2019one}\cite{zhou2019d}\cite{clerigues2020acute}\cite{qi2019x}\cite{pedemonte2018detection}. In terms of brain-like neural nets,  Kubilius et al. demonstrated that better anatomical alignment to the brain of deep learning networks could achieve high performance in image recognition tasks \cite{kubilius2019brain}. Alekseev et al. proposed an image recognition system by using a learnable Gabor Layer inside the deep convolutional neural network \cite{alekseev2019gabornet}, since Gabor filter is able to effectively extract spacial frequency structures from images and represent texture features.

\subsection{Modeling}
\begin{figure}[h!]
\begin{center}
\includegraphics[width=10cm]{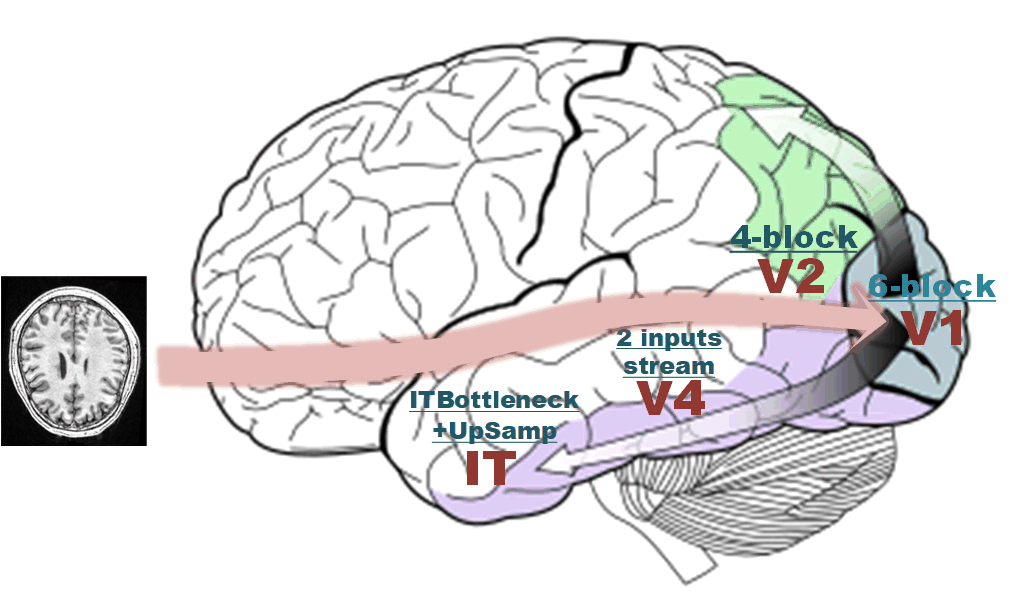}
\end{center}
\caption{Computational modeling of the visual cortex anatomy (The ventral stream begins with V1, goes through visual area V2, then through visual area V4, and to the inferior temporal cortex)}\label{fig:arch}
\end{figure}

This work proposes to construct a visual cortex anatomy alike model to simulate the process of identifying the stroke lesion using human vision system. The model is simulated as a deep neural network follows the anatomical structure of the human visual cortex. The visual cortex processes visual information inside human brain by transmitting visual information into two primary pathways, known as the ventral stream and the dorsal stream \cite{mishkin1983object}. The ventral stream is associated with object representation, form recognition, and storage of long-time memory. The dorsal stream is associated with motion and representation of object locations. This work only considers the ventral steam and models the ventral anatomical structure considering the characteristics of this specific segmentation task. An overview of the computational segmentation modeling of visual cortex is illustrated in \autoref{fig:arch}.

\begin{figure}[h!]
\begin{center}
\includegraphics[width=10cm]{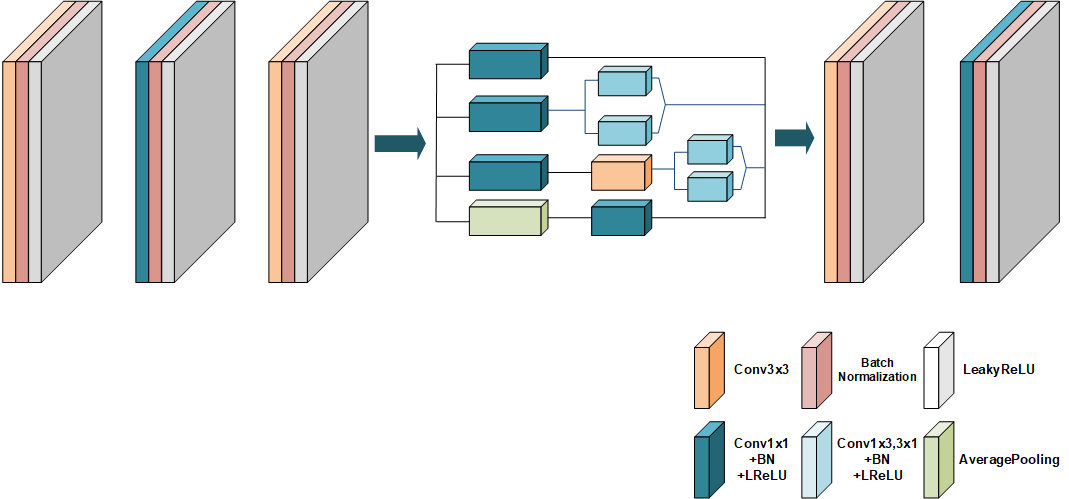}
\end{center}
\caption{V1 Block}\label{fig:V1}
\end{figure}

The primary visual cortex (V1) consists of six functionally distinct layers and layer 4 is further divided into 4 separate layers \cite{hubel1972laminar}. The receptive field of neurons in V1 is typically smaller compared to other visual cortex microscopic regions. In view of the structure and functionality of the primary visual cortex, the V1 area is modeled as a customized sets of layers. It is constructed with 6 convolutional blocks and the fourth block is further divided into 4 parallel layers \autoref{fig:V1}.

\begin{figure}[h!]
\begin{center}
\includegraphics[width=10cm]{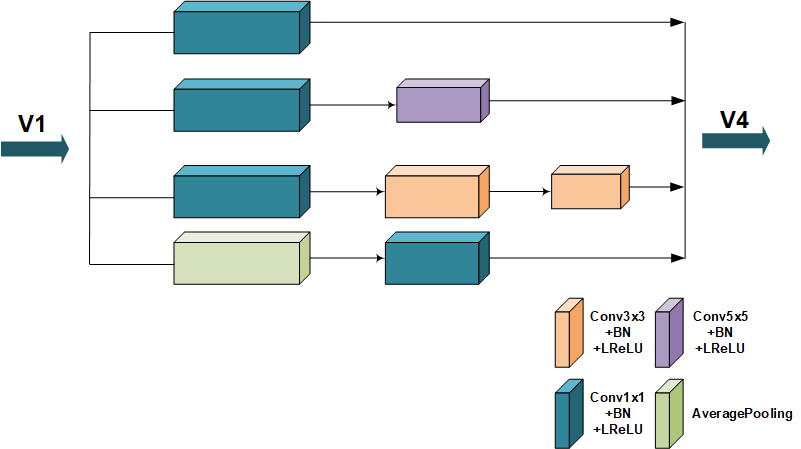}
\end{center}
\caption{V2 Block}\label{fig:V2}
\end{figure}

Visual area V2 is the second major area in the visual cortex. It receives feedforward connections from V1 while sending feedbacks to V1, and further sends connections to V3, V4, and V5 \cite{schwarzkopf2011surface}. V2 contains a dorsal and ventral representation in the left and the right hemispheres and cells are tuned to simple properties including orientation, spatial frequency, and color, like V1 structure. In view of the common properties between V1 and V2 and the anatomical structure of V2 area, V2 is modeled as a 4 parallel-layers module \autoref{fig:V2}.

\begin{figure}[h!]
\begin{center}
\includegraphics[width=10cm]{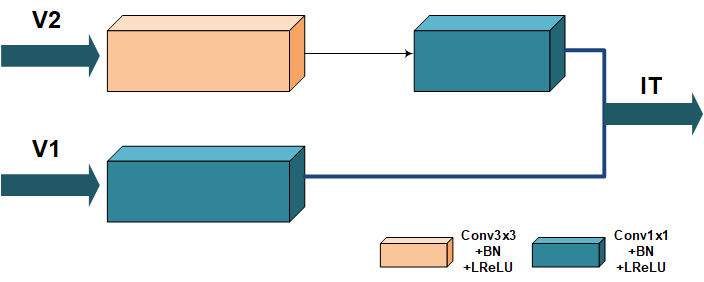}
\end{center}
\caption{V4 Block}\label{fig:V4}
\end{figure}

Visual area V4 is located anterior to V2 and posterior to the posterior inferotemporal area (PIT). It receives strong feedforward input from V2 and also receives direct input from V1 \cite{goddard2011color}. V4 is also tuned for orientation, spatial frequency, and color and is the first area in the ventral stream to show strong attentional modulation. V4 is further tuned for object features of intermediate complexity. V4 is computationally modeled as a simple stack of layers but accepts two inputs which represents the intermediate input from V2 and direct input from V1 \autoref{fig:V4}.

\begin{figure}[h!]
\begin{center}
\includegraphics[width=10cm]{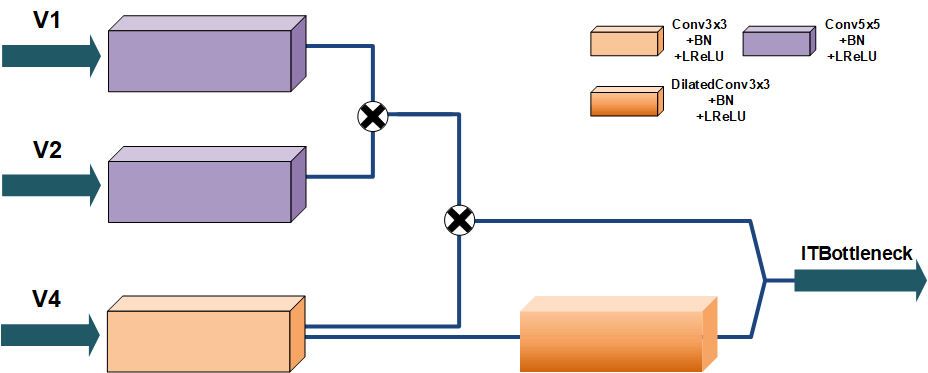}
\end{center}
\caption{IT Bottleneck}\label{fig:IT}
\end{figure}

The IT cortex in humans is located to a specific region of the human temporal lobe \cite{chelazzi1993neural}\cite{logothetis1995shape}. It processes visual stimuli of objects and is involved with memory and memory recall to identify specific objects \cite{miyashita1993inferior}. It processes visual stimuli from V1, V2, V3, and V4 regions of the occipital lobe \cite{gross1992representation} by identifying the object based on the color and form and comparing that processed information to stored memories of objects to identify specific object \cite{kolb2001introduction}. IT cortex is designed to amplify the features calculated by previous areas.
IT area is modeled as a more complicated block of layers compared to other areas \autoref{fig:IT}, first it accepts three separate inputs corresponding to V1, V2, and V4 intermediate outputs. After transforming these three inputs through custom convolutional blocks, the transformed $v_{1}$ and $v_{2}$ will be merged and the merged results will further be merged with the transformed $v_{4}$ inputs into $v_{m}$. Finally, the transformed $v_{4}$ will go through a dilated convolutional block and then be concatenated with the merged results $v_{m}$. This intermediate results will serve as the IT bottleneck and further be followed by 4 UpSampling blocks to output the final prediction. The kernel size of IT module is typically larger than previous areas.

\subsection{Training Strategy}
Stroke lesion varies in size and shape, which makes it hard to predict. Also in a supervised context, the target lesion mask samples are usually far less than the normal samples which makes it even hard to train. This paper utilizes the EMLLoss function proposed by \cite{zhou2019d} and combines it with the Binary Cross Entropy Loss. The EMLLoss is constructed with Focal Loss and Dice Loss. Focal Loss is proposed by \cite{lin2017focal} to alleviate the extreme data imbalance issue. Dice Loss is derived from the Dice Similarity Coefficient which measures the similarity between two sample sets \cite{milletari2016v}.

\begin{equation}
DSC=\frac{2TP}{2TP+FP+FN}\label{eq:dsc}
\end{equation}

\begin{equation}
\begin{aligned}
Loss&=EMLLoss+BCELoss\\
&=\frac{-\alpha_{t}(1-p_{t})^{\gamma}log(p_{t})}{N} + log(1 - DSC) + BCELoss\label{eq:loss}
\end{aligned}
\end{equation}

\section{Experiments}
\subsection{Implementation}
\subsubsection{Neural Nets}
The V1 module consists of a six-layer basic convolution block. Each basic convolution block is formed by sequentially lining up a convolutional layer, a BatchNormalization layer, and a LeakyReLU activation layer (Conv-BN-LReLU). The kernel size of the first, third, and fifth convolutional layer of V1 is set to 3 with a stride of 1. The kernel size of the second, sixth layer is set to 1 with a stride of 1. The fourth layer is an InceptionE alike module with different in-out channels setup and the V2 module is an InceptionA alike module with different in-out channels setup. 
The V4 module consists of two branches, the first branch stacks two Conv-BN-LReLU blocks with convolutional kernel size of 3 and 1 respectively; the second branch is a single Conv-BN-LReLU block with convolutional kernel size of 1.
The IT module consists a IT bottleneck block and a UpSampling block. The IT bottleneck is constructed with four Conv-BN-LReLU blocks with kernel size and stride setting to 5(stride=2), 5(2), 3(1), 3(padding=2, stride=1, dilation=2), respectively. Each of them corresponds to a transform of the corresponding inputs $v_{1}$, $v_{2}$, $v_{4}$, and $v_{m}$. Four UpSampling blocks are appended after to restore the intermediate feature maps into the original resolution. Skip connections between each stage are also implemented in this task.

\begin{figure}[h!]
\begin{center}
\includegraphics[width=0.9\textwidth]{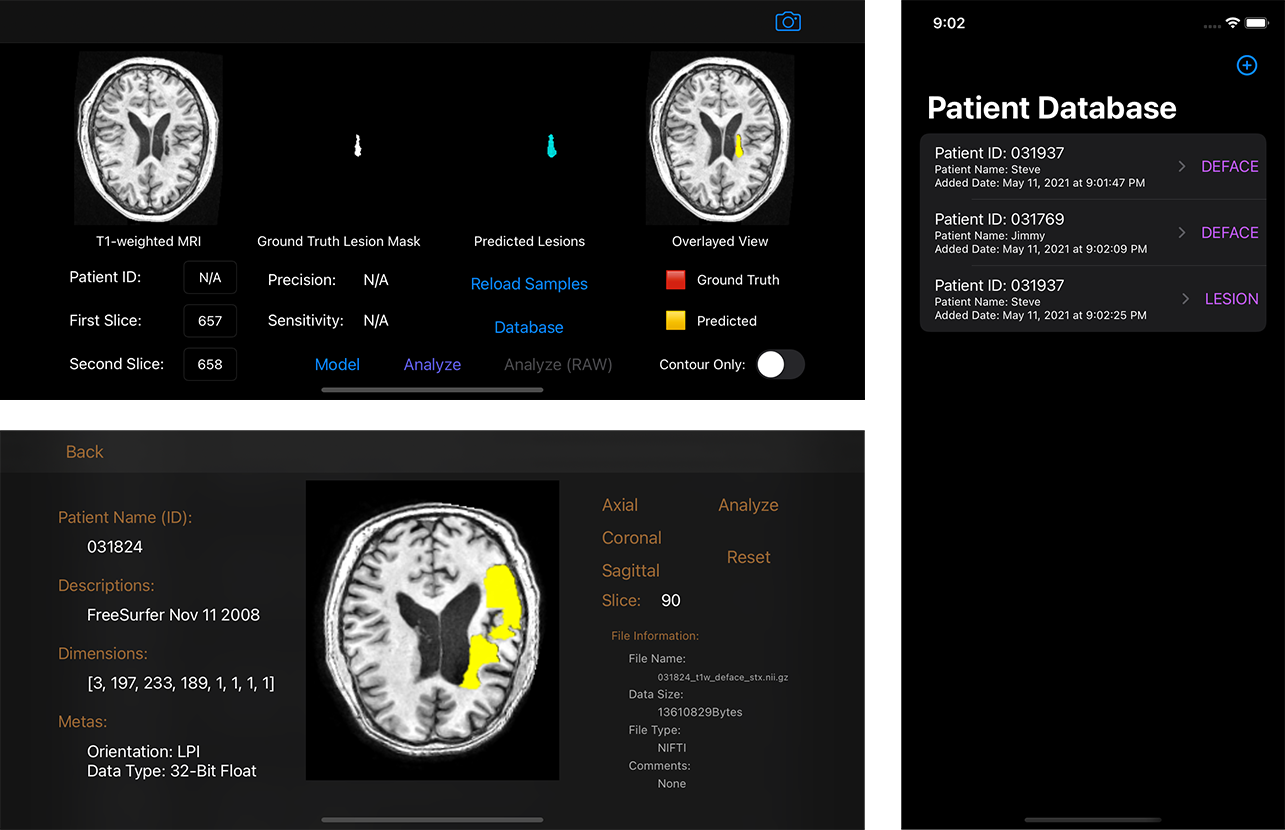}
\end{center}
\caption{Prototype system on iOS}\label{fig:ss}
\end{figure}

The SGD optimizer is used with an initial learning rate of 0.001, a weight decay of 1e-8 and a momentum of 0.9. The batch size is set to 8 due to the limitation of the GPU memory. All network code are written in Python using PyTorch framework and the model was trained using one NVIDIA GTX2080Ti GPU. A prototype stroke imaging application on iOS is also developed to facilitate the usage of the model. The stroke imaging application has been tested on an iPhone 11 and an iPhone 7 plus model and performs well. Details could be seen in \autoref{fig:ss}.

\subsubsection{Dataset}
The Anatomical Tracings of Lesions After Stroke (ATLAS) dataset is used in this work \cite{liew2018large}. The ATLAS dataset contains 229 T1-weighted MRI images and each volume contains 189 axial slices and each slice is of size 233x197. All axial image slices are resized into size of 224x192 and a split of 9:1 of all the preprocessed axial slices is used during the training process. 

\subsection{Analysis}
\subsubsection{Metrics}
In a disease context, true positive (TP) stands for correctly diagnosed as diseased; false positive (FP) stands for incorrectly diagnosed as diseased; true negative (TN) stands for correctly diagnosed as not diseased; false negative (FN) stands for incorrectly diagnosed as not diseased.
Jaccard Index (IoU) and Dice similarity coefficient (F1 score) are both used to measure the similarity of two sets, the mathematical representations can be written using the definition of TP, FP, and FN:

\begin{equation}
F_{1}=\frac{2 * precision * recall}{precision + recall}=\frac{TP}{TP + \frac{1}{2}(FP + FN)}\label{eq:f1}
\end{equation}

\begin{equation}
J(A, B)=\frac{A \cap B}{A\cup B}=\frac{A \cap B}{|A| + |B| - |A \cap B|}\label{eq:iou}
\end{equation}

Sensitivity (recall, true positive rate), refers to the proportion of positive cases that are correctly identified:

\begin{equation}
SENSITIVITY=\frac{TP}{TP+FN}\label{eq:recall}
\end{equation}

Precision (positive predictive value), measures the fraction of positive cases among the retrieved instances: 

\begin{equation}
PRECISION=\frac{TP}{TP+FP}\label{eq:precision}
\end{equation}

The Hausdorff distance is the greatest of all the distances from a point in one set to the closest point in the other set. It measures the distance between two subsets of a certain metric space:

\begin{equation}
d_{H}(X,Y)=\mathrm{max}\{\mathop{\mathrm{sup}}\limits_{x\in X} d(x,Y), \mathop{\mathrm{sup}}\limits_{y\in Y} d(X,y)\}\label{eq:hd}
\end{equation}

\subsubsection{Results}
\begin{figure}[h!]
  \centering
  \includegraphics[width=0.65\textwidth]{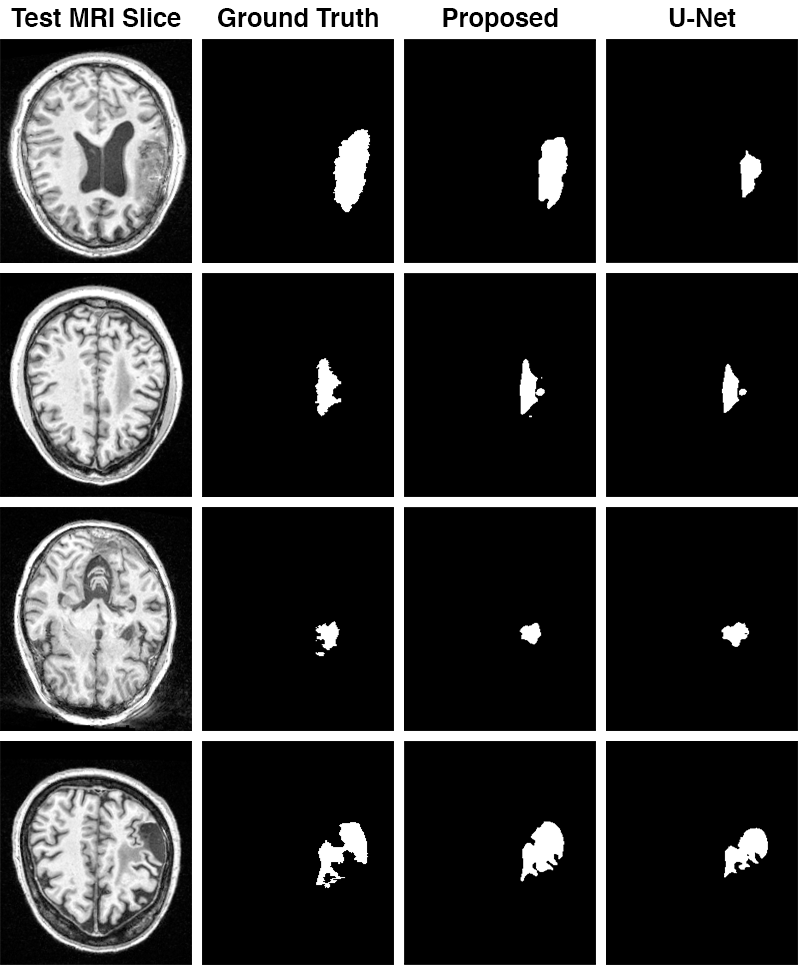}
  \caption{Segmentation examples on the ATLAS dataset}
  \label{fig:examples}
\end{figure}

Some of the tested images from the ATLAS dataset are shown in \autoref{fig:examples}. State-of-the-art performance is not the goal of this work as the intention is to model the network follows the visual cortex anatomy so as to demonstrate the neuroscience intuition. Still, the proposed model can perform comparably to the de-facto standard U-Net \autoref{tab:metrics}.

\begin{table}[h!]
  \centering
  \caption{Evaluation metrics on the ATLAS dataset (Slice-to-Slice)}
  \begin{tabular}{l@{~~~}l@{~~~}l@{~~~}l@{~~~}l@{~~~}l@{~~~}ll}
    \toprule
    Method       &  DSC & IoU & HD & Sensitivity & Precision & F1 Score \\
    \midrule
    U-Net 	      & 0.8806 & 0.2935 & 22.0066 & 0.3508 & 0.4457 & 0.3687 \\
    Proposed   & 0.8449   & \textbf{0.3379} & 32.9180 & \textbf{0.4296} & \textbf{0.5349} & \textbf{0.4454} \\
    \bottomrule\\[-0.8cm]
  \end{tabular}
  \label{tab:metrics}
\end{table}

\section{Discussions}
Stroke lesion segmentation is a very challenging task considering its nature. This paper proposes a computational model which utilizes neural networks to resemble the anatomical structure of the visual cortex. This research is conducted primarily by demonstrating how neural networks should be designed from a neuroscience point of view, which could bring a more explanatory model for doctors and physicians thus putting more trust into the computational models, which could further facilitate the usage of the models and be of real-world assistance. However, this work is just a preliminary research and only considers part of the anatomical structure of the visual cortex. To further improve the model and research on this area, more complex brain anatomical connections and functionalities of the brain areas should be considered into. Sophisticated net structures should also be meticulously designed.

\section*{Acknowledgments}
This work was supported by the Postgraduate Research \& Practice Innovation Program of Jiangsu Province (Grant KYCX20\_0934). The author would like to thank Dr. David S. Liebeskind and Dr. Fabien Scalzo for their inspirations and thoughtful insights, and Dr. Xingming Sun for providing the infrastructure resources for the experiments to be conducted smoothly.

\bibliographystyle{splncs04}
\bibliography{arxiv}
\end{document}